\documentclass[11pt]{article}
\usepackage{draft}
\usepackage{cite}
\usepackage{cancel}
\usepackage{mathrsfs}
\usepackage{enumitem}
\usepackage{xcolor}
\usepackage{caption}  
\usepackage{graphicx} 
\usepackage{float} 
\usepackage{cite}
\usepackage{relsize}
\usepackage{physics}
\usepackage{psfrag}
\usepackage{cancel}
\usepackage{array}
\usepackage{amssymb}
\usepackage{amsmath}
\usepackage[compat=1.1.0]{tikz-feynman}
\usepackage{amsthm}
\usepackage{float}
\usepackage{tikz}
\usetikzlibrary{patterns}
\usetikzlibrary{mindmap,decorations.pathmorphing,backgrounds,positioning,fit}
\usetikzlibrary{decorations.markings}
\usepackage{tikz,lipsum,lmodern}
\usepackage[most]{tcolorbox}
\usepackage{hyperref}
\usepackage{xcolor}
\usepackage{tikz}
\usetikzlibrary{decorations.pathmorphing,patterns}
\usepackage{amsmath}
\usepackage{amssymb}
\usepackage{float}
\usepackage{fixmath}
\usepackage{physics}
\usepackage{slashed}
\usepackage{graphicx}
\usepackage{mathrsfs}
\usepackage{amsbsy}
\usepackage{subfig}
\usepackage{multirow}
\usepackage{hyperref}
\usepackage{bbm}

\usepackage{physics}
\usepackage{amsmath}
\usepackage{tikz}
\usepackage{mathdots}
\usepackage{yhmath}
\usepackage{cancel}
\usepackage{color}
\usepackage{siunitx}
\usepackage{array}
\usepackage{multirow}
\usepackage{amssymb}
\usepackage{gensymb}
\usepackage{tabularx}
\usepackage{extarrows}
\usepackage{booktabs}
\usetikzlibrary{fadings}
\usetikzlibrary{patterns}
\usetikzlibrary{shadows.blur}
\usetikzlibrary{shapes}
\usetikzlibrary{decorations.markings}
\tikzset{mid arrow/.style={postaction={decorate}, decoration={markings, mark=at position 0.5 with {\arrow{Triangle[angle=45:.21cm 1.5]}}}}}
\tikzset{
pattern size/.store in=\mcSize, 
pattern size = 5pt,
pattern thickness/.store in=\mcThickness, 
pattern thickness = 0.3pt,
pattern radius/.store in=\mcRadius, 
pattern radius = 1pt}
\makeatletter
\pgfutil@ifundefined{pgf@pattern@name@_cu8hrzqwm}{
\pgfdeclarepatternformonly[\mcThickness,\mcSize]{_cu8hrzqwm}
{\pgfqpoint{0pt}{0pt}}
{\pgfpoint{\mcSize+\mcThickness}{\mcSize+\mcThickness}}
{\pgfpoint{\mcSize}{\mcSize}}
{
\pgfsetcolor{\tikz@pattern@color}
\pgfsetlinewidth{\mcThickness}
\pgfpathmoveto{\pgfqpoint{0pt}{0pt}}
\pgfpathlineto{\pgfpoint{\mcSize+\mcThickness}{\mcSize+\mcThickness}}
\pgfusepath{stroke}
}}
\makeatother
\definecolor{twilightlavender}{rgb}{0.54, 0.29, 0.42}
\definecolor{richmaroon}{rgb}{0.69, 0.19, 0.38}
\definecolor{forestgreen(web)}{rgb}{0.13, 0.55, 0.13}
\definecolor{lava}{rgb}{0.81, 0.06, 0.13}
\hypersetup{
	breaklinks,
	colorlinks,
	citecolor=forestgreen(web),
	filecolor=richmaroon,
	linkcolor=lava,
	urlcolor=twilightlavender
}

\usepackage{cleveref}

\crefformat{section}{\S#2#1#3} 
\crefformat{subsection}{\S#2#1#3}
\crefformat{subsubsection}{\S#2#1#3}

\usepackage{verbatim}
 
\usepackage{color}

\title{Waveform, memory and classical soft scalar theorems}
\affiliation[a]{Yau Mathematical Sciences Center (YMSC), Tsinghua University, Beijing 100084, China}
\affiliation[b]{Department of Theoretical Physics (DTP),
Tata Institute of Fundamental Research (TIFR), Homi Bhabha Road, Mumbai 400005, India}
\usepackage{orcidlink}
\author[a,\orcidlink{0000-0002-4535-3198}]{Sarthak Duary,}\emailAdd{sarthakduary@tsinghua.edu.cn}
\author[b,\orcidlink{0000-0002-5553-7003}]{and Pabitra Ray}\emailAdd{raypabitra96@gmail.com}

\abstract{We examine a scattering process in which a set of particles come together, interact through a long-range massless scalar force like dilaton mediated and then disperse. Using worldline formalism, we compute the trajectories of the scattered particles and derive an infinite series of subleading terms in the late-time and early-time expansion of the scalar waveform. We study the scalar waveform, the scalar memory term and classical soft scalar theorems. We ignore gravitational interactions and discuss the challenges if we include it.}
\begin{document}
\maketitle

\section{Introduction}
We consider a scattering process in which a set of particles approach each other from asymptotic infinity, interact via a massless scalar interaction which is inherently long-range and then disperse to future null infinity. During this interaction, classical scalar radiation is emitted, which is a direct consequence of the massless nature of the mediating field. This scalar field can be naturally interpreted as a dilaton, a fundamental degree of freedom appearing in the low-energy effective actions of string theory. The dilaton is a massless scalar particle that occurs together with the graviton and the Kalb-Ramond antisymmetric tensor in the massless sector of both the closed bosonic string theory and the NS-NS (Neveu-Schwarz) sector of closed superstring theory. In string theory, the dilaton plays a crucial role: it determines the string coupling constant and governs the strength of gravitational and gauge interactions in the effective spacetime description. In particular, in closed bosonic string theory and superstring theories in ten dimensions, the dilaton arises alongside the graviton and antisymmetric tensor fields as part of the massless spectrum of the closed string. Accordingly, the interaction we consider may be understood as a dilaton-mediated force, analogous to gravity, but scalar in nature. The resulting scalar radiation emitted during the scattering process is thus not only a signature of the long-range interaction but also encodes information about the asymptotic symmetries and soft structure associated with the dilaton. \vspace{0.2cm}\\
Classical soft theorems impose universal constraints on the structure of electromagnetic and gravitational waveforms in the low-frequency (or long-wavelength) regime. These constraints govern the nonanalytic behavior of the waveform features such as poles or branch cuts in the frequency domain-which cannot be captured by a simple Taylor expansion around zero frequency. By virtue of the Fourier transform, these low-frequency characteristics correspond to distinct features in the time-domain signal at asymptotically early or late times. \vspace{0.2cm}\\
One of the most celebrated manifestations of this infrared structure is the \emph{gravitational memory effect} \cite{Zeldovich:1974gvh,Christodoulou:1991cr,Wiseman:1991ss,Thorne:1992sdb}. This phenomenon describes a permanent change in the spacetime geometry-specifically, a lasting displacement in the transverse spatial metric-at future null infinity, induced by the passage of gravitational radiation. In the frequency domain, this translates into a leading-order singularity proportional to $ \omega^{-1} $, where $ \omega $ is the angular frequency of the emitted radiation \cite{Weinberg:1964ew,Weinberg:1965nx,Strominger:2014pwa}. \vspace{0.2cm}\\
The physical intuition behind this effect lies in the nature of soft graviton emissions. In particular, the dominant contributions to the waveform in the limit of vanishing frequency arise from two distinct sources\,:
\begin{enumerate}
    \item Emissions associated with the initial and final massive states involved in the scattering process, which give rise to what is known as \emph{linear memory}.
    
    \item Graviton emissions from other emitted gravitons themselves, resulting in a nonlinear contribution referred to as \emph{nonlinear memory}.
\end{enumerate}
These soft theorems, together with the associated memory effects and an underlying infinite-dimensional symmetry group at null infinity, form what is now commonly referred to as \emph{infrared triangle} \cite{Strominger:2013jfa,He:2014laa,He:2014cra,Strominger:2014pwa,Campiglia:2015qka,Kapec:2015ena,Strominger:2017zoo}. This triangle encodes deep structural relationships between the infrared behavior of massless fields, the symmetries of asymptotically flat spacetimes, and the conservation laws that constrain physical observables. It has become a cornerstone in understanding the interplay between classical and quantum aspects of gravity and gauge theories in the infrared limit. \vspace{0.2cm}\\
We study a classical scattering process in which particles interact via a long-range \textit{massless scalar field}, emitting scalar radiation. This field can be interpreted as a dilaton, an essential field in the low-energy effective description of string theory. 
In closed bosonic string theory, the massless spectrum includes three key fields the graviton $ g_{\mu\nu} $, which governs spacetime geometry, the Kalb-Ramond field $ B_{\mu\nu} $, an antisymmetric tensor field, the dilaton $ \phi $, a scalar field.
These combine into the low-energy effective action of string theory (in the Einstein frame)
\begin{equation}
    S = \frac{1}{2\kappa^2} \int d^{D}x \sqrt{-g} \left[ R - \frac{1}{2} \partial_\mu \phi\partial^\mu\phi - \frac{1}{12} H_{\mu\nu\rho} H^{\mu\nu\rho} + \cdots \right]~,
\end{equation}
 where $ H_{\mu\nu\rho} \equiv \partial_{[\mu} B_{\nu\rho]} $ is the field strength of the Kalb--Ramond field. The dilaton plays a central role in determining the coupling constants of the theory
\begin{equation}
g_{\mathrm{s}} = \mathrm{e}^{\expval{\phi}}~,
\end{equation}
where $ \expval{\phi} $ is the vacuum expectation value\,(VEV) of the dilaton. Therefore, when we talk about a massless scalar mediating an interaction, we essentially describe dilaton-mediated forces, forces that arise from the exchange of a quantum of the dilaton field.\vspace{0.4cm}\\
While the dilaton is a fundamental scalar in string theory, it is not the only such scalar. A particularly interesting example is the QCD axion, introduced to solve the strong CP problem in QCD. Like the dilaton, the axion is a \textit{light\,(or massless in the high-temperature limit)} scalar field with derivative couplings, leading to long-range forces under certain conditions. The QCD axion arises as a pseudo-Nambu-Goldstone boson associated with spontaneous breaking of the Peccei-Quinn symmetry. Axions also appear generically in string compactifications, where they descend from higher-dimensional $ p $-form fields reduced on nontrivial cycles of the internal manifold. These so-called \textit{string axions} often remain massless or acquire small masses due to non-perturbative effects, making them viable candidates for dark matter and mediators of new long-range forces.\vspace{0.4cm}\\
The waveform of this radiation depends on the detailed acceleration of the particles during the scattering. However, at late-time and early-time, when the particles are nearly free, the situation simplifies since the only forces acting on them are the long-range massless scalar. In \cite{Saha:2019tub, Laddha:2018myi, Laddha:2018vbn, Sahoo:2018lxl}, it was established that the asymptotic behavior of the radiation waveform observed at future null infinity (\(\mathscr{I}^+\)) exhibits a universal structure determined by the properties of the scattering particles. Specifically, if \( u \) denotes the retarded time measured by a detector placed at \( \mathscr{I}^+ \), with \( u=0 \) corresponding to the moment when the signal peak reaches the detector, then the coefficients of the constant term, the term \( 1/u \) and the term \( u^{-2} \ln \abs{u} \) in the waveform at large positive and large negative \( u \) are completely determined by the momenta and charges of the incoming and outgoing particles. Recently, in \cite{Karan:2025ndk} for electromagnetic interaction, it was generalized by demonstrating that a similar structure holds for a broader class of terms appearing in the late-time  and early-time expansions of the waveform. Specifically in \cite{Karan:2025ndk}, it was shown that the coefficients of the terms of the form \( u^{-r-1} (\ln \abs{u})^r \) for all \( r \geq 0 \) can also be expressed purely in terms of the momenta and charges of the scattered particles.\vspace{0.4cm}\\ 
There is a connection between the quantum soft theorems governing $\mathcal{S}$-matrix elements and their classical counterparts. In particular, previous studies have established a direct link between the late-time structure of classical radiation waveform and the quantum soft theorems in gauge theories. For instance, the electromagnetic waveform at orders \( u^{-1} \) and \( u^{-2} \ln \abs{u} \) has been shown to correspond to the quantum soft photon theorem at orders \( \ln \omega \) and \( \omega (\ln \omega)^2 \), respectively, for soft photon energy \( \omega \). This connection was derived in \cite{Sahoo:2018lxl} by analyzing loop quantum electrodynamics (QED) $\mathcal{S}$-matrices, see also \cite{Sen:2024bax}.\vspace{0.4cm}\\ 
Motivated by this interplay between quantum and classical soft theorems in gauge theory, we initiate a systematic exploration of the scalar waveform and the classical soft theorems governing massless scalar interactions using worldline formalism. While the study of soft theorems for gauge bosons and gravitons has seen significant progress, the analogous structure for scalar interactions remains less understood. By extending the classical soft theorem framework to massless scalar fields, we aim to uncover universal features of scalar radiation and establish a foundation for future investigations into the quantum-classical correspondence in scalar theories. Specifically, an analysis of \((r+1)\)-loop QED $\mathcal{S}$-matrices is expected to give the soft photon theorem of order \( \omega^r (\ln \omega)^{r+1} \), which, in the appropriate classical limit, should correspond to the electromagnetic waveform of order \( u^{-r-1} (\ln u)^r \) as derived in \cite{Karan:2025ndk}.\vspace{0.4cm}\\
The work \cite{Karan:2025ndk} introduces a systematic prescription for deriving these classical waveforms by first calculating the trajectories of scattered particles using trajectory coefficients. We apply the prescription to massless scalar interactions, which is a linear theory. We give a complete study of the prescription in which the approach can be applied because of the linearity of the theory. In this paper, we study massless scalar interactions and systematically compute an infinite hierarchy of subleading terms in the asymptotic expansions of the scalar waveform, both at late-time and early-time. While the massless scalar case is instructive, we recognize the limitations of the approach when attempting to generalize to nonlinear theories such as gravity; for details, we refer to the conclusions section. We need novel techniques for handling nonlinear theories like gravity. \vspace{0.4cm}\\
In \cite{Campiglia:2017dpg}, soft scalar theorems and their formulation within the framework of asymptotic symmetries were examined. In \cite{Campiglia:2017dpg}, it was shown that the leading-order contributions of the soft scalar theorem depend solely on the momenta and scalar charges of the incoming and outgoing particles. However, beyond these leading contributions, the waveform contains an infinite sequence of subleading corrections that refine our understanding of the asymptotic structure of radiation. \vspace{0.4cm}\\
In this paper, we systematically derive this entire hierarchy of subleading corrections to the scalar waveform using worldline formalism. So far, discussions of soft scalar theorems \cite{Campiglia:2017dpg, Biswas:2022lsj, Derda:2024jvo, Briceno:2025ivl} have been largely confined to tree-level scattering amplitudes. This is in contrast to other soft theorems, such as those for gauge bosons and gravitons, which have been studied more extensively beyond the tree level. A key challenge in extending soft-scalar theorems to loop-level calculations arises from the fact that the massless scalar field under consideration typically acquires a mass due to quantum corrections. As a result, taking a well-defined soft limit becomes problematic, implying that the associated asymptotic symmetries are broken at the quantum level.\vspace{0.4cm}\\
For the special case of two-particle scattering, we obtain a fully resummed expression for scalar radiation. In addition, we derive a resummed form of the waveform that remains valid in both the asymptotic future and the past, capturing the complete late-time and early-time behavior of the emitted radiation. We ignore gravitational interactions. We give a detailed discussion of the key challenges if we incorporate gravitational interactions.
\subsection*{Summary of the set-up.}
\begin{figure}[h]
\begin{center}
\begin{tikzpicture}
\draw[line width=0.4mm] (0,3) -- (-3,0) -- (0,-3) -- (3,0) -- (0,3) -- (-3,0);
\draw[pattern= north east lines,line width=0.4mm] (0,0) circle [radius=1cm];

\draw[mid arrow,line width=0.4mm] (-0.57,0.81) to [out=120,in=-110](0,3);

\draw[mid arrow,line width=0.4mm]     (0.5,0.86) to [out=60,in=-70] (0,3);

\draw[mid arrow,line width=0.4mm]  (0,-3)   to [out=80,in=-70] (0.5,-0.86) ;

\draw[mid arrow,line width=0.4mm]    (0,-3) to [out=120,in=-120] (-0.5,-0.86) ;

\fill (0,1.3) circle [radius=0.03cm];
\fill (-0.18,1.29) circle [radius=0.03cm];
\fill (-0.36,1.25) circle [radius=0.03cm];

 \node[above right] at (1.5,1.5) {$\mathscr{I}^+$};
\node[above left] at (-1.5,1.5) {$\mathscr{I}^+$};

\fill (-0.18,-1.29) circle [radius=0.03cm];
\fill (0,-1.3) circle [radius=0.03cm];
\fill (0.18,-1.29) circle [radius=0.03cm];

\draw[-{Triangle[angle=45:.21cm 1.5]},line width=0.4mm,densely dotted,]  (0.8,0.8) -- (1.2,1.2) ;
\draw[line width=0.4mm,densely dotted]  (0.8,0.8) -- (1.38,1.38) ;

\draw[-{Triangle[angle=45:.21cm 1.5]},line width=0.4mm,densely dotted]  (-0.8,0.8) -- (-1.2,1.2) ;
\draw[line width=0.4mm,densely dotted,]  (-0.8,0.8) -- (-1.38,1.38) ;

\node[below right] at (1.5,-1.5) {$\mathscr{I}^-$};
\node[below left] at (-1.5,-1.5) {$\mathscr{I}^-$};

\node[right] at (3,0) {$i^0$};
\node[left] at (-3,0) {$i^0$};

\node[above] at (0,3) {$i^+$};
\node[below] at (0,-3) {$i^-$};

\end{tikzpicture}
\caption{The scattering process.}
    \label{30}    
\end{center}
\end{figure}
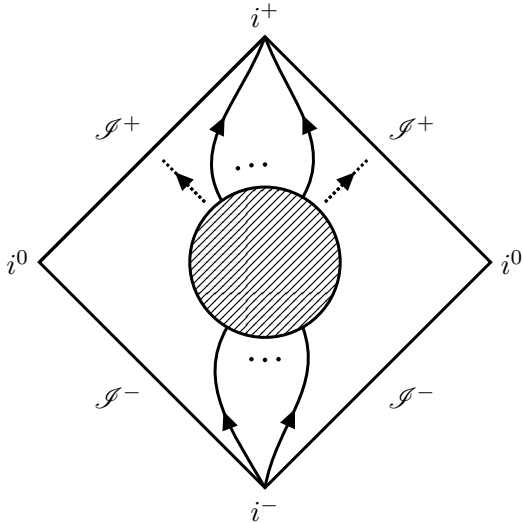

Now, we explain our set-up as shown in \figurename\,[\ref{30}]. 
We consider a system where $N'$ incoming particles, each with masses $\{m'_{a}\}$ and momenta $\{p'_{a}\}$, undergo a process of collision, fusion, and subsequent fragmentation, resulting in $N$ outgoing particles with masses $\{m_a\}$ and momenta $\{p_a\}$. Let \(\vb{x}\) denote the spatial coordinates of a detector placed in the future null infinity\,(\(\mathscr{I}^+\)) in the \figurename\,[\ref{30}] to capture the scattered scalar waveform. We introduce the following definitions 
\begin{equation}
 R\equiv  \abs{\vb{x}}~, \quad \hat{\vb{n}}\equiv\frac{\vb{x}}{R}~, \quad n\equiv(1,\hat{\vb{n}})~, \quad u\equiv x^0 - R~,
\end{equation}
where $u$ denotes the retarded time and we adopt natural units with $c=1$. The peak of the scalar signal arrives at the detector in $u \sim 1$.   
We now examine the waveform in the limit of large positive $u$, which is determined by the asymptotic trajectories of the outgoing particles. The relevant part of the trajectory \cite{Karan:2025ndk} for the $a$-th outgoing particle is given by  
\begin{equation}\label{10}
X^\mu_a(\tau_a) = \frac{p^\mu_a}{m_a} \tau_a + \left(C^{(a)\mu}_1 \ln\tau_a + \mathcal{O}(1) \right) - \sum_{r=1}^{\infty} \left( C^{(a)\mu}_{r+1} \left( \frac{\ln\tau_a}{\tau_a} \right)^r + \mathcal{O}(\tau_a^{-r} (\ln\tau_a)^{r-1}) \right) + \cdots~~,   
\end{equation}
where $\tau_a$ is some parameter along the worldline of the $a$-th particle, $p^\mu_a$ is its momentum, $m_a$ is its mass, and $C^{(a)\mu}_{r+1}$ are constants.
These terms will be essential to identify the terms of the scalar waveform $u^{-r-1} (\ln |u|)^r \quad \text{for all } r \geq 0$. Our goal is to preserve all terms of order $u^{-r-1} (\ln u)^r$ in the waveform. To achieve this, we formally introduce a large parameter $\lambda$ and assume that both $u$ and $\ln u$ scale with $\lambda$. As we shall see later, when calculating the waveform in the detector at retarded time $u$, the parameter $\tau_a$ corresponding to the trajectory of the $a$-th particle is also of order $u$. Consequently, $\tau_a$ scale as $\lambda$ and their logarithms, $\ln \tau_a \simeq \ln u$.\vspace{0.4cm}\\ 
For the trajectory of the incoming particles, we replace $\tau_a$ with $-\tau_a$ and \(\ln \tau_a\) with \(\ln \abs{\tau_a}\). As a result, \( C^{(a)\mu}_{r+1} \) gets an extra factor \( (-1)^r \), and outgoing momenta, masses, and charges are replaced with incoming ones. 

\subsection*{Organization of the paper} 
The organization of the paper is as follows. In \cref{sec1}, we compute particle trajectories. In \cref{subsec1}, we compute the trajectory coefficients for two-body scattering. Next, in \cref{trajprobapp}, we compute the trajectory under the probe approximation that simplifies our analysis. In \cref{sec3}, we compute the scalar waveform, and also determine the scalar memory term in \cref{memsc}. Next, in \cref{subsec3.1} and \cref{subsec3.3}, we focus on the computation of a scalar waveform specifically in the context of two-body scattering scenarios and analyze our result for a purely elastic process. Next, in \cref{subsec3.2}, we compute the scalar waveform under the probe approximation. Finally, in \cref{conc}, we conclude the paper by summarizing our key findings and providing suggestions for future directions.

\section{Particles trajectories}
\label{sec1}
In this \cref{sec1}, we compute the particle trajectories. The equations of motion that govern the dynamics of the trajectory of the particles $X^\mu_a(\tau_a)$ and the scalar field $\phi_{(b)}(x)$ are given by
\begin{equation}
\label{eom}
 \begin{split}
&m_a\frac{d^2X^\mu_a(\tau_a)}{d\tau_a^2}=q_a\sum_{b\neq a}\partial^\mu \phi_{(b)}(X_a(\tau_a))\\
&\square \phi_{(b)}(x)=-J_{(b)}(x)~.
 \end{split}   
\end{equation}
This equation describes the motion of a particle \( a \) with mass \( m_a \) under the influence of a scalar field \( \phi_{(b)}(x) \). The force on the particle \( a \) arises from the gradient of the field \( \phi_{(b)}(x) \) sourced by other particles \( b \). The coupling constant \( q_a \) represents the charge of the particle with respect to the scalar field $\phi(x)$.
This is the equation of motion for the massless scalar field \(\phi_{(b)}(x)\), following a standard Klein-Gordon like form with a source term \( J_{(b)}(x) \). The source \( J_{(b)}(x) \) is usually related to the charges \( q_b \) and the worldline of the particle \( b \), typically of the form
\begin{equation}
J_{(b)}(x) = q_b \int d\tau_b\,\delta^{(4)}\big(x - X_b(\tau_b)\big)~,
\end{equation}
which represents a point source moving along its worldline.
If the field is massless, the interaction is long-range (like electromagnetism but scalar in nature). The derivation of the particle equation of motion is given in the Appendix \ref{pfdvv}.\\
The solution $\phi_{(b)}(x)$ is given by
\begin{equation}
\phi_{(b)}(x)=\frac{q_b}{2\pi}\int d\tau_b~\delta_+\big(-(x-X_b(\tau_b))^2\big)~.
\end{equation}
The function $\delta_+$ represents the standard Dirac delta function, which is multiplied by the Heaviside theta function $H\big(x^0 - X_b^0(\tau_b)\big)$. Therefore, we need to select the zero argument for which $x^0 > X_b^0(\tau_b)$. Taking the first derivative, we get
\begin{equation}
\begin{split}
&\partial^\mu \phi_{(b)}(x)\\
&=-\frac{q_b}{2\pi}\int d\tau_b \, \frac{\big(x^\mu-X^\mu_{b}(\tau_b)\big)}{\big(x-X_{b}(\tau_b)\big)\cdot\frac{dX_b(\tau_b)}{d\tau_b}}\left[\pdv{\tau_b}\delta_+\big(-(x-X_b(\tau_b))^2\big) \right]~~~~~~~~~~~~~~~~\\&=-\frac{q_b}{2\pi}\int d\tau_b\,\delta_+\big(-(x-X_b(\tau_b))^2\big)\Bigg[\frac{1}{\big(x-X_{b}(\tau_b)\big)\cdot\frac{dX_b(\tau_b)}{d\tau_b}}\frac{dX^\mu_b(\tau_b)}{d\tau_b}\\&~~~~~~~~~~~~~~~~+\frac{\big(x^\mu-X^\mu_b(\tau_b)\big)}{\left(\big(x-X_{b}(\tau_b)\big)\cdot\frac{dX_b(\tau_b)}{d\tau_b}\right)^2}\left\{\big(x-X_b(\tau_b)\big)\cdot\frac{d^2X_b(\tau_b)}{d\tau_b^2}-\left(\frac{dX_b(\tau_b)}{d\tau_b}\right)^2\right\}\Bigg]~.
\end{split}
\end{equation}
To handle the delta function, we use the identity for the derivative of the delta function
\begin{equation}
\begin{split}
\pdv{x_\mu}\delta_+\big(-(x-X_b(\tau_b))^2\big)&=-2\big(x^\mu-X^\mu_{b}(\tau_b)\big)\delta'_+\big(-(x-X_b(\tau_b))^2\big)\\&=-\frac{\big(x^\mu-X^\mu_{b}(\tau_b)\big)}{\big(x-X_{b}(\tau_b)\big)\cdot\frac{dX_b(\tau_b)}{d\tau_b}}\left[\pdv{\tau_b}\delta_+\big(-(x-X_b(\tau_b))^2\big) \right]~. 
\end{split}
\end{equation}
To compute the right-hand side of the equations of motion \eqref{eom} for particle \( a \), we need to evaluate the field strength at its location, which requires computing \( \partial^\mu \phi_{(b)}(x) \). This, in turn, involves determining the parameter \( \tau_b \) of the particle \( b \) in terms of the parameter \( \tau_a \) of the particle \( a \). We find this by solving the constraint equation,  
\begin{equation}
\big(X_a(\tau_a) - X_b(\tau_b)\big)^2 = 0~,
\end{equation}
which is imposed by the delta function, ensuring that the interaction occurs along the light cone. Now, we must carefully decide which terms in the trajectory expansions of \( X^\mu_a(\tau_a) \) and \( X^\mu_b(\tau_b) \) should be retained and which can be neglected. Since our objective is to capture all terms of order \( u^{-r-1} (\ln u)^r \) in the scalar waveform, we introduce a formal expansion parameter \( \lambda \) and assume that both \( u \) and \( \ln u \) scale as \( \lambda \). This scaling assumption helps us to systematically track the relative importance of different terms in the trajectory expansion.\vspace{0.4cm}\\ 
A crucial observation is that, for computing the waveform in a detector at retarded time \( u \), the relevant parameters \( \tau_a \) and \( \tau_b \) are also of order \( u \). Consequently, we deduce that \( \tau_a \sim \lambda \) and \( \tau_b \sim \lambda \), while their logarithms, \( \ln \tau_a \simeq \ln u \) and \( \ln \tau_b \simeq \ln u \), also scale as \( \lambda \). Examining the expansion of \( X^\mu_a(\tau_a) \), we find that the terms proportional to \( \tau_a \) and \( \ln \tau_a \) are of order \( \lambda \), while the coefficients \( C_r^{(a)\mu} \) for \( r \geq 2 \) remain of order \( \lambda^0 \). On the other hand, any terms neglected on the \( X^\mu_a(\tau_a) \) scale as \( \lambda^{-p} \) for \( p \geq 1 \), making their contribution negligible for our analysis.\vspace{0.4cm}\\
Therefore, when solving equation \( \big( X_a(\tau_a) - X_b(\tau_b)\big)^2 = 0 \) to express \( \tau_b \) in terms of \( \tau_a \), we retain only the leading contributions from the expansions of \( X^\mu_a(\tau_a) \) and \( X^\mu_b(\tau_b) \). Specifically, we keep the first two terms in their respective expansions while discarding higher-order corrections, ensuring that our calculations remain consistent with the desired order of accuracy in the scalar waveform.\vspace{0.4cm}\\
In order to solve the constraint equation $\big(X_a(\tau_a)-X_b(\tau_b)\big)^2=0$~, we will use the first two terms of \eqref{10} which give us 
\begin{equation}\label{3}
X^\mu_c(\tau_c)=\frac{p^\mu_c}{m_c}\tau_c+C^{(c)\mu}_1\ln\tau_c+\mathcal{O}(\lambda^0)\quad\textmd{where}~~c=a,b~~,
\end{equation}
and equation $\big(X_a(\tau_a)-X_b(\tau_b)\big)^2=0$ reduces to
\begin{equation}
\begin{split}
 &\tau_a^2+\tau_b^2+2\frac{p_a\cdot p_b}{m_am_b}\tau_a\tau_b+2\frac{p_a\cdot\left( C^{(b)}_1-C^{(a)}_1\right)}{m_a}\tau_a\ln\tau_a-2\frac{p_b\cdot\left( C^{(b)}_1-C^{(a)}_1\right)}{m_b}\tau_b\ln\tau_a\\&~~~~-\left( C^{(b)}_1-C^{(a)}_1\right)^2(\ln\tau_a)^2+\mathcal{O}(\lambda)=0~.
\end{split}   
\end{equation}
The solution for $\tau_b$ gives us
\begin{equation}
\tau_b=\tau_{ba}^\star+\mathcal{O}(\lambda^0)~,
\end{equation}
and $\tau_{ba}^\star$ is given by
\begin{equation}
\begin{split}
\tau_{ba}^\star&=-\frac{p_a\cdot p_b}{m_am_b}\tau_a+\frac{p_b\cdot\left(C_1^{(b)}-C^{(a)}_1\right)}{m_b}\ln\tau_a-\left[-\tau_a^2-2\frac{p_a\cdot\left(C_1^{(b)}-C^{(a)}_1\right)}{m_a}\tau_a\ln\tau_a\right.\\&~~~~+\left. \left(C_1^{(b)}-C^{(a)}_1\right)^2(\ln\tau_a)^2+\left(\frac{p_a\cdot p_b}{m_am_b}\tau_a-\frac{p_b\cdot\left(C_1^{(b)}-C^{(a)}_1\right)}{m_b}\ln\tau_a\right)^2\right]^{\frac{1}{2}}~,
\end{split}   
\end{equation}
where we have chosen the lowest value of $\tau_b$ due to the Heaviside step function $H\big(X^0_a(\tau_a)-X^0_b(\tau_b)\big)$ in the retarded solution of the Green function.\vspace{0.4cm}\\
We will substitute $x^\mu=X^\mu_a(\tau_a)$ and the delta function constraint becomes
\begin{equation}
\begin{split}
\delta_+\big(-(X_a(\tau_a)-X_b(\tau_b))^2\big)=\frac{1}{2\abs{\big(X_a(\tau_a)-X_b(\tau_b)\big)\cdot\frac{dX_b(\tau_b)}{d\tau_b}}}_{\tau_b=\tau^\star_{ba}}\delta(\tau_b-\tau_{ba}^\star)~.
\end{split}    
\end{equation}
Performing the delta function integral, we get
the final result as
\begin{equation}
\begin{split}
&\partial^\mu \phi_{(b)}(X_a(\tau_a))\\&=-\frac{q_b}{4\pi}\frac{1}{\abs{\big(X_a(\tau_a)-X_b(\tau_b)\big)\cdot\frac{dX_b(\tau_b)}{d\tau_b}}}_{\tau_b=\tau^\star_{ba}}
\\&~~~~\times\Bigg[\frac{1}{\big(X_a(\tau_a)-X_{b}(\tau_b)\big)\cdot\frac{dX_b(\tau_b)}{d\tau_b}}\frac{dX^\mu_b(\tau_b)}{d\tau_b}+\frac{\big(X^\mu_a(\tau_a)-X^\mu_b(\tau_b)\big)}{\left(\big(X_a(\tau_a)-X_{b}(\tau_b)\big)\cdot\frac{dX_b(\tau_b)}{d\tau_b}\right)^2}\\&~~~~~~~~~~~~~~~~~~~~~~~~\times\left\{\big(X_a(\tau_a)-X_b(\tau_b)\big)\cdot\frac{d^2X_b(\tau_b)}{d\tau_b^2}-\left(\frac{dX_b(\tau_b)}{d\tau_b}\right)^2\right\}\Bigg]_{\tau_b=\tau^\star_{ba}}.
\end{split}
\end{equation}
Substituting the expressions for $X^\mu_{c}(\tau_c)$, we get
\begin{equation}
\begin{split}
&\partial^\mu \phi_{(b)}(X_a(\tau_a))\\&=-\frac{q_b}{4\pi}\frac{1}{\tau_a^2}\frac{m_a^2m_b}{\left[(p_a\cdot p_b)^2-m_a^2m_b^2\right]^{\frac{3}{2}}}\\&~~~~\times\left[1-2m_a\frac{m^2_bp_a\cdot\left(C^{(b)}_1-C^{(a)}_1\right)+(p_a\cdot p_b)p_b\cdot\left(C^{(b)}_1-C^{(a)}_1\right)}{(p_a\cdot p_b)^2-m^2_a m^2_b}\xi_a\right.\\&~~~~~~~~~~~~~~~~~~~~~~~~~~~~~~~~~~~\left.+m_a^2\frac{\left(p_b\cdot\left(C^{(b)}_1-C^{(a)}_1\right)\right)^2+m^2_b\left(C^{(b)}_1-C^{(a)}_1\right)^2}{(p_a\cdot p_b)^2-m^2_a m^2_b}\xi_a^2\right]^{-\frac{3}{2}}\\&~~~~\times\Bigg[p^\mu_b\left(p_a\cdot p_b\right)+m_b^2p^\mu_a+m_ap_b\cdot\left(C^{(a)}_1-C^{(b)}_1\right)p^\mu_b\xi_a+m_am^2_b\left(C^{(a)\mu}_1-C^{(b)\mu}_1\right)\xi_a\Bigg]+\mathcal{O}(\lambda^{-3})~,
\end{split}    
\end{equation}
where $\xi_a\equiv\frac{\ln\tau_a}{\tau_a}$. We compute the first derivative and the second derivative of $X^\mu_a(\tau_a)$ and compare it with \eqref{3}. The derivatives are
\begin{equation}
\begin{split}
&\frac{d X^\mu_a(\tau_a)}{d\tau_a}=\frac{p^\mu_a}{m_a}+\mathcal{O}(\lambda^{-1})\\& \frac{d^2 X^\mu_a(\tau_a)}{d\tau^2_a}= -C_1^{(a)\mu}\frac{1}{\tau_a^2}-\sum_{r=1}^{\infty}r(r+1)C^{(a)\mu}_{r+1}\frac{(\ln\tau_a)^r}{\tau_a^{r+2}}+\mathcal{O}(\lambda^{-3})~.
\end{split}   
\end{equation}
\subsection{Trajectory coefficients}
We get the equation for the trajectory coefficients
\begin{equation}
\label{imp}
\begin{split}
&C_1^{(a)\mu}+\sum_{r=1}^{\infty}r(r+1)C^{(a)\mu}_{r+1}\xi_a^r\\&\simeq \sum_{b\neq a}\frac{q_aq_b}{4\pi}\frac{m_am_b}{\left[(p_a\cdot p_b)^2-m_a^2m_b^2\right]^{\frac{3}{2}}}\\&~~~~\times\Bigg[p^\mu_b\left(p_a\cdot p_b\right)+m_b^2p^\mu_a-m_ap_b\cdot\left(C^{(b)}_1-C^{(a)}_1\right)p^\mu_b\xi_a-m_am^2_b\left(C^{(b)\mu}_1-C^{(a)\mu}_1\right)\xi_a\Bigg] 
\\&~~~~\times\left[1-2m_a\frac{m^2_bp_a\cdot\left(C^{(b)}_1-C^{(a)}_1\right)+(p_a\cdot p_b)p_b\cdot\left(C^{(b)}_1-C^{(a)}_1\right)}{(p_a\cdot p_b)^2-m^2_a m^2_b}\xi_a\right.\\&~~~~~~~~~~~~~~~~~~~~~~~~~~~~~~~~~~~~~\left.+m_a^2\frac{\left(p_b\cdot\left(C^{(b)}_1-C^{(a)}_1\right)\right)^2+m^2_b\left(C^{(b)}_1-C^{(a)}_1\right)^2}{(p_a\cdot p_b)^2-m^2_a m^2_b}\xi_a^2\right]^{-\frac{3}{2}}~.
\end{split}   
\end{equation}
To solve for the trajectory coefficient \( C_1^{(a) \mu} \), we set \(\xi_a = 0\), and it simplifies to
\begin{equation}
\begin{split}
C_1^{(a) \mu} &=\sum_{b\neq a}\frac{q_aq_b}{4\pi}\frac{m_am_b}{\left[(p_a\cdot p_b)^2-m_a^2m_b^2\right]^{\frac{3}{2}}} \Big[p^\mu_b\left(p_a\cdot p_b\right)+m_b^2p^\mu_a\Big]~.
\end{split}
\end{equation}
For $r \geq 1$, iteratively solving, we get the trajectory coefficients $C^{(a)\mu}_{r+1}$ 
\begin{equation}
\label{trajcoeffnn}
\begin{split}
C^{(a)\mu}_{r+1}&=\frac{m_a^{r}}{2^{r}r(r+1)}\sum_{b\neq a}\frac{m_am_b}{\left[(p_a\cdot p_b)^2-m_a^2m_b^2\right]^{\frac{3}{2}}}\Bigg[\left(p^\mu_b\left(p_a\cdot p_b\right)+m_b^2p^\mu_a\right)\\&~~~~\times\sum_{k=\lfloor\frac{r}{2}\rfloor}^{r}\frac{(-1)^k(2k+1)!}{k!(2k-r)!(r-k)!}\frac{1}{\left[(p_a\cdot p_b)^2-m_a^2m_b^2\right]^k}
\\&~~~~\times\left(m^2_b p_a\cdot\left(C^{(a)}_1-C^{(b)}_1\right)+(p_a\cdot p_b)p_b\cdot\left(C^{(a)}_1-C^{(b)}_1\right)\right)^{2k-r}\\&~~~~~~~~~~~~~~~~~~~~~~~~~~~~~\times\left(\left(p_b\cdot\left(C^{(a)}_1-C^{(b)}_1\right)\right)^2+m^2_b\left(C^{(a)}_1-C^{(b)}_1\right)^2\right)^{r-k}
\\&~~~~+2\left(p_b\cdot\left(C^{(a)}_1-C^{(b)}_1\right)p^\mu_b + m^2_b\left(C^{(a)\mu}_1-C^{(b)\mu}_1\right)\right)\\&~~~~\times\sum_{k=\lfloor\frac{r-1}{2}\rfloor}^{r-1}\frac{(-1)^k(2k+1)!}{k!(2k+1-r)!(r-k-1)!}\frac{1}{\left[(p_a\cdot p_b)^2-m_a^2m_b^2\right]^k}
\\&~~~~~~~~~~~\times\left(m^2_b p_a\cdot\left(C^{(a)}_1-C^{(b)}_1\right)+(p_a\cdot p_b)p_b\cdot\left(C^{(a)}_1-C^{(b)}_1\right)\right)^{2k+1-r}\\&~~~~~~~~~~~~~~~~~~~~~~~~~~~~~~~~~~~\times\left(\left(p_b\cdot\left(C^{(a)}_1-C^{(b)}_1\right)\right)^2+m^2_b\left(C^{(a)}_1-C^{(b)}_1\right)^2\right)^{r-k-1}\Bigg]~.
\end{split}    
\end{equation}
Here, in deriving \eqref{trajcoeffnn} we have used the following identity to extract the trajectory coefficients $C^{(a)\mu}_{r+1}$ 
\begin{equation}
\big(1+A\xi_a+B\xi_a^2\big)^{-\frac{3}{2}}= \sum_{k=0}^\infty \sum_{s=0}^k (-1)^k \frac{(2k+1)!}{2^{2k} k! s!(k-s)!} A^s B^{k-s} \xi_a^{2k-s}~.    
\end{equation}
In \cite{Karan:2025ndk} it has been observed that the trajectory coefficients in electromagnetic interactions satisfy condition \( p_a \cdot C_1^{(a)} = 0 \). We now give a proof of this. This condition arises because of the relation  
\begin{equation}
\frac{dX_a(\tau_a)}{d\tau_a} \cdot \frac{d^2X_a(\tau_a)}{d\tau^2_a} = 0~,\quad\Rightarrow~p_a\cdot C^{(a)}_1+\sum_{r=1}^{\infty} r(r+1) p_a\cdot C^{(a)}_{r+1}\xi_a^r=0~.
\end{equation}
This is a consequence of the Lorentz force law. However, in the case of scalar interactions, similar conditions on the trajectory coefficients do not hold. This marks a key difference between the two types of interactions.\vspace{0.4cm}\\
The trajectory analysis at early times follows a similar approach, with a few key differences. First, the terms that include \(\ln \tau_a\) are replaced by \(\ln \abs{\tau_a}\).  Another difference comes from the absolute value in the denominator of \(\partial^\mu \phi_{(b)}(X_a(\tau_a))\), which causes \(\abs{\tau_a}\) to become \(-\tau_a\). This results in a sign change on the right-hand side of the equation that determines the trajectory coefficients. However, this sign change can be corrected by redefining \[C_r^{(a)\mu} \to (-1)^r C_r^{(a)\mu}~, \quad \xi_a \to -\xi_a~.\]
As a result, the final expressions of the trajectory coefficients \( C^{(a)\mu}_{r+1} \) have an additional factor of \( (-1)^r \), and the outgoing momenta and charges must be replaced with the incoming ones.

\subsection{Trajectory for the two-body scattering}
\label{subsec1}
In this \cref{subsec1}, we compute the trajectory of two-body scattering. We simplify and get 
\begin{equation}
\begin{split}
C_1^{(a) \mu} &=\frac{q_aq_b}{4\pi}\frac{m_am_b}{\left[(p_a\cdot p_b)^2-m_a^2m_b^2\right]^{\frac{3}{2}}} \Big[p^\mu_b\left(p_a\cdot p_b\right)+m_b^2p^\mu_a\Big]~, \quad \text{for } b \neq a
\end{split}
\end{equation}
which means
\begin{equation}
p_a\cdot C^{(b)}_1= 0~,\quad p_a\cdot C^{(a)}_1=p_b\cdot C^{(b)}_1=\frac{q_aq_b}{4\pi}\frac{m_am_b}{\left[(p_a\cdot p_b)^2-m_a^2m_b^2\right]^{\frac{1}{2}}}~, \quad \text{for } b \neq a~.  
\end{equation}
We also find that conditions \( p_a \cdot C_1^{(b)} = 0 \)\,(for $b\neq a$) hold, which are newly observed constraints. These conditions naturally emerge from the trajectories in two-body scattering. In contrast to electromagnetic interactions, where the generic trajectory coefficient satisfies \( p_a \cdot C_1^{(a)} = 0 \), we see a notable difference here: instead of the momentum of particle \( a \) imposing the condition on the trajectory coefficient \( C_1^{(a)} \), it is the momentum of particle \( b \) that does so. This difference is striking and we believe that it represents a potentially significant new observation in the context of scalar interactions. \vspace{0.2cm}\\
We can simplify the following expression as a perfect square 
\begin{equation}
\begin{split}
&1-2m_a\frac{m^2_bp_a\cdot\left(C^{(b)}_1-C^{(a)}_1\right)+(p_a\cdot p_b)p_b\cdot\left(C^{(b)}_1-C^{(a)}_1\right)}{(p_a\cdot p_b)^2-m^2_a m^2_b}\xi_a\\&~~~~~~~~~~~~~~~~~~~~~~~~~~~~~~~~~~~~~~~~~~+m_a^2\frac{\left(p_b\cdot\left(C^{(b)}_1-C^{(a)}_1\right)\right)^2+m^2_b\left(C^{(b)}_1-C^{(a)}_1\right)^2}{(p_a\cdot p_b)^2-m^2_a m^2_b}\xi_a^2\\&=\left[1-m_a\frac{m^2_bp_a\cdot\left(C^{(b)}_1-C^{(a)}_1\right)+(p_a\cdot p_b)p_b\cdot\left(C^{(b)}_1-C^{(a)}_1\right)}{(p_a\cdot p_b)^2-m^2_a m^2_b}\xi_a\right]^2
\\&=\left[1-\frac{q_aq_b}{4\pi}\frac{m_am_bp_b\cdot(p_a+p_b)}{\left[(p_a\cdot p_b)^2-m^2_a m^2_b\right]^{\frac{3}{2}}}m_a\xi_a\right]^2~.
\end{split}
\end{equation}
Using this simplification, the final result for the trajectory coefficient $C_{r+1}^{(a)\mu}$ is given by
\begin{equation}
C_{r+1}^{(a)\mu} = \frac{m_a^r}{r} \left( \frac{q_a q_b}{4\pi} \frac{m_a m_b}{\left[ (p_a \cdot p_b)^2 - m_a^2 m_b^2 \right]^{\frac{3}{2}}} \right)^{r+1} \left[ p_b \cdot (p_a + p_b) \right]^r \left( p_b^\mu (p_a \cdot p_b) + m_b^2 p_a^\mu \right)~, \quad \text{for } b \neq a~.
\end{equation}

\subsection{Trajectory in the probe approximation}
\label{trajprobapp}
In this \cref{trajprobapp}, we analyze the trajectory using the probe approximation. We examine a scattering process that involves \( N \) probe particles and a heavy scatterer. The probe particles are specified by charge \( \{q_a\} \), mass \( \{m_a\} \), and outgoing momentum \( \{p_a\} \) \( \forall a = 1,2, \dots, N \) that interact with the massive scatterer of charge \( Q \) and mass \( M \) at rest. The probe limit is characterized by the conditions \( M \gg m_a \), \( Q \gg q_a \) and \( \frac{Q}{M} \ll \frac{q_a}{m_a} \) \( \forall a = 1,2, \dots, N \). The requirement \( \frac{Q}{M} \ll \frac{q_a}{m_a} \) ensures that the heavy scatterer remains effectively stationary under the scalar influence of the probe particles, allowing us to disregard its scalar radiation.\vspace{0.2cm}\\
The momenta of the heavy scatterer and the probe particles are given by
\begin{equation}
    P = \left(M, \vb{0} \right)~, \quad p_a = \frac{m_a}{\sqrt{1 - \beta_a^2}} \left(1, \mathbold{\beta}_a \right)~, \quad \forall a = 1,2, \dots, N~.
\end{equation}
Within this probe approximation, the trajectory coefficients for the probe particles are solely influenced by the long-range scalar force exerted by the heavy scatterer.
The leading correction to the trajectory coefficient $C_1^{(a)}$, simplifies to
\begin{equation}
\label{prapp}
    C_1^{(a)} \simeq  \frac{q_a Q}{4\pi} \frac{(1 - \beta_a^2)}{m_a \beta_a^3} \left(0, \mathbold{\beta}_a \right)~, \quad \forall a = 1,2, \dots, N~.
\end{equation}
The trajectory coefficient $C_{r+1}^{(a)\mu}$ simplifies in the probe limit as
\begin{equation}
\label{prapp2}
\begin{split}
C_{r+1}^{(a)} &\simeq
\frac{(-1)^r}{r} \frac{q_a Q}{4\pi} \frac{(1 - \beta^2_a)}{m_a \beta_a^3} \left( \frac{q_a Q}{4\pi} \frac{(1 - \beta_a^2)^{\frac{3}{2}}}{m_a \beta_a^3} \right)^r\left( 0, \mathbold{\beta}_a \right)~.
\end{split}
\end{equation}
It should be noted that the temporal component vanishes in \eqref{prapp} and \eqref{prapp2}, which is in stark contrast to the electromagnetic case recently studied in \cite{Karan:2025ndk}. When the field is massless, the interaction is long-range, similar to electromagnetism but with a scalar nature instead. This suggests a different behavior for the forces involved, with potential implications for studying scalar interactions in the context of classical field theories. The absence of temporal component in the trajectory coefficients $C^{(a)\mu}_r$ is consistent with the momentum constraints $P\cdot C^{(a)}_r=0$ for all $r\ge 1$. In addition, the coefficients do not depend on the mass $M$. This is due to the heavy nature of the scatterer.
\section{Generic scalar waveform}
\label{sec3}
In this \cref{sec3}, we calculate the generic scalar waveform. After determining the trajectories of the particles, we can compute the scalar waveform at the detector. This analysis follows the same approach as that of the recent article \cite{Karan:2025ndk}. The solution scalar waveform $\phi(x)$ is given by
\begin{equation}
\label{solphib}
\phi(x)=\sum_{a}\frac{q_a}{2\pi}\int d\tau_a~\delta_+\big(-(x-X_a(\tau_a))^2\big)~.
\end{equation}
From the trajectories of the outgoing particles, we can apply \eqref{solphib} to calculate the scalar waveform. 
We parameterize the position \( x \) of the detector as \( x^\mu \equiv (u + R, R \hat{\vb{n}}) = (u, \vb{0}) + R n \). For large values of \( R \) and fixed \( u \), the delta function constraint leads to the following expression
\begin{equation}
\big(x - X_a(\tau_a)\big)^2 = 0~,\quad \Rightarrow \quad -2uR - 2R \, n \cdot X_a(\tau_a) + \mathcal{O}(R^0) = 0~.
\end{equation}
The leading order term in \( R \) yields
\begin{equation}
- u = \frac{n \cdot p_a}{m_a} \tau_a + n \cdot C_1^{(a)} \ln \tau_a - \sum_{r=1}^{\infty} n \cdot C_{r+1}^{(a)} \left( \frac{\ln \tau_a}{\tau_a} \right)^r + \mathcal{O}(\lambda^{-1})~.
\end{equation}
We observe that the first two terms on the right-hand side are of order \( \lambda \), while the remaining terms are of order \( \lambda^0 \). Thus, we retain only the first two terms and solve for \( \tau_a \)
\begin{equation}
\tau_a = \tau_a^{\text{sol}}~, \quad \tau_a^{\text{sol}} \simeq -\frac{u \, m_a}{n \cdot p_a} \left( 1 + n \cdot C_1^{(a)} w \right)~, \quad w \equiv \frac{\ln u}{u}~,
\end{equation}

\begin{equation}
\Rightarrow \quad \xi_a^{\text{sol}} \equiv \frac{\ln \tau_a^{\text{sol}}}{\tau_a^{\text{sol}}} \simeq -\frac{n \cdot p_a}{m_a} w \left( 1 + n \cdot C_1^{(a)} w \right)^{-1}~.
\end{equation}
Thus, in the limit of large \( R \) and large retarded time \( u \), the delta function constraint reduces to
\begin{equation}
\delta_+ \left( - (x - X_a(\tau_a))^2 \right) \simeq \frac{1}{2R} \frac{\delta (\tau_a - \tau_a^{\text{sol}})}{\abs{n \cdot \frac{dX_a(\tau_a)}{d\tau_a}}}_{\tau_a=\tau_a^{\text{sol}}}~.
\end{equation}
We use 
\begin{equation}
\frac{dX^\mu_a(\tau_a)}{d\tau_a}\simeq\frac{p_a^\mu}{m_a}+C_1^{(a)\mu}\frac{1}{\tau_a}+\sum_{r=1}^{\infty}rC_{r+1}^{(a)\mu}\frac{(\ln\tau_a)^r}{\tau_a^{r+1}}+\mathcal{O}(\lambda^{-2})~.
\end{equation}
We observe that the first term on the right-hand side is of order \(\lambda^0\), whereas the remaining terms are of order \(\lambda^{-1}\). Since we will compute terms in the scalar waveform up to the first subleading order in the \(\lambda^{-1}\) expansion, it is essential to retain all terms.
So, the late-time scalar radiation waveform is given by 
\begin{equation}
\begin{split}
 &\phi(u,R\hat{\vb{n}})\\&\simeq-\frac{1}{4\pi R}\sum_{a}\frac{q_am_a}{n\cdot p_a}\int d\tau_a\,\delta\big(\tau_a-\tau_a^{\textmd{sol}}\big)\frac{1}{1+\frac{m_a}{n\cdot p_a}\frac{n\cdot C_1^{(a)}}{\tau_a}+\frac{m_a}{n\cdot p_a}\sum_{r=1}^{\infty}rn\cdot C_{r+1}^{(a)}\frac{(\ln\tau_a)^r}{\tau_a^{r+1}}+\mathcal{O}(\lambda^{-2})}~,   
\end{split}
\end{equation}
using \( n \cdot p_a \) negative. We evaluate the integral over \( \tau_a \) and retain terms up to order \( \lambda^{-1} \), yielding
\begin{equation}
\begin{split}
&\phi(u,R\hat{\vb{n}})\\&\simeq-\frac{1}{4\pi R}\sum_{a}\frac{q_am_a}{n\cdot p_a}+\frac{1}{4\pi R}\sum_{a}\frac{1}{\tau_a^{\textmd{sol}}}q_a\left(\frac{m_a}{n\cdot p_a}\right)^2\Bigg[n\cdot C_1^{(a)}+\sum_{r=1}^{\infty}rn\cdot C_{r+1}^{(a)}\left(\xi_a^{\text{sol}}\right)^r\Bigg]+\mathcal{O}(\lambda^{-2})~.
\end{split}
\end{equation}
Substituting $\tau_a^{\textmd{sol}}$ and $\xi_a^{\textmd{sol}}$, we have the expression of the final scalar waveform
\begin{equation}
\begin{split}
\phi(u,R\hat{\vb{n}})&\simeq-\frac{1}{4\pi R}\sum_{a}\frac{q_am_a}{n\cdot p_a}\\&~~~~~~~~~+\frac{1}{4\pi R}\sum_{r=1}^{\infty}(-1)^{r}\frac{(\ln u)^{r-1}}{u^r}\sum_{a}\frac{q_am_a}{n\cdot p_a}\Bigg[\left(n\cdot C_1^{(a)}\right)^r\\&~~~~~~~~~~~~~~+\sum_{k=1}^{r-1}\frac{(r-1)!}{(k-1)!(r-k-1)!}n\cdot C_{k+1}^{(a)}\left(\frac{n\cdot p_a}{m_a}\right)^k\left(n\cdot C_1^{(a)}\right)^{r-k-1}\Bigg]+\mathcal{O}(\lambda^{-2})~.
\end{split}
\end{equation}
The scalar waveform result for large negative \( u \) is derived by substituting the outgoing charges and momenta with the incoming charges and momenta, replacing $u$ with $-u$, and \( \ln u \) with \( \ln |u| \) also incorporating an additional factor of \( (-1)^r \) in the expression for \( C_r^{(a)\mu} \).
\subsection{Scalar memory term}
\label{memsc}
The $u$ independent term in the scalar waveform is given by $-\frac{1}{4\pi R}\sum_{a}\frac{q_am_a}{n\cdot p_a}$. This term is not the scalar memory term. In the study of scalar radiation, the scalar waveform does not fade even as the retarded time \( u \) approaches positive infinity (\( u \to +\infty \)). This persistence indicates that the scalar field retains a ``memory'' of past events, a concept known as the \textit{scalar memory effect}. To simplify our analysis and make the behavior of the scalar field more transparent, we have added a constant $+\frac{1}{4\pi R}\sum_{a}\frac{q'_{a}m'_{a}}{n\cdot p'_{a}}$ to the scalar waveform. This adjustment ensures that the waveform vanishes as \( u \to -\infty \). By doing so, we are effectively setting a reference point in the distant past where the scalar field is zero. This is what we refer to as the \textit{observer's frame}. In this frame, an observer measures changes in the scalar field starting from this calibrated baseline. Now, the redefined scalar waveform for positive $u$ becomes
\begin{equation}
\begin{split}
\phi(u,R\hat{\vb{n}})&\simeq-\frac{1}{4\pi R}\sum_{a}\frac{q_am_a}{n\cdot p_a}+\frac{1}{4\pi R}\sum_{a}\frac{q'_am'_a}{n\cdot p'_a}\\&~~~~~~~~~+\frac{1}{4\pi R}\sum_{r=1}^{\infty}(-1)^{r}\frac{(\ln u)^{r-1}}{u^r}\sum_{a}\frac{q_am_a}{n\cdot p_a}\Bigg[\left(n\cdot C_1^{(a)}\right)^r\\&~~~~~~~~~~~~~~+\sum_{k=1}^{r-1}\frac{(r-1)!}{(k-1)!(r-k-1)!}n\cdot C_{k+1}^{(a)}\left(\frac{n\cdot p_a}{m_a}\right)^k\left(n\cdot C_1^{(a)}\right)^{r-k-1}\Bigg]+\mathcal{O}(\lambda^{-2})~.
\end{split}
\end{equation}    
The desired \textit{scalar memory term} is given by
\begin{equation}
 -\frac{1}{4\pi R}\sum_{a}\frac{q_am_a}{n\cdot p_a}+\frac{1}{4\pi R}\sum_{a}\frac{q'_am'_a}{n\cdot p'_a}~.   
\end{equation}
We can think of it as setting our stopwatch to zero before starting a race. By establishing a common starting point, any subsequent measurements accurately reflect the changes from that initial moment. Similarly, redefining the waveform to vanish at \( u \to -\infty \), we can precisely track how the scalar field evolves due to radiation emitted by sources.\footnote{Unlike transient oscillations that dissipate over time, these memory effects leave a lasting imprint on the field. This is analogous to the \textit{gravitational memory effect} in General Relativity, where spacetime itself retains a permanent deformation after gravitational waves pass through.} The waveform in negative $u$ is given by
\begin{equation}
 \begin{split}
\phi(u,R\hat{\vb{n}})&\simeq\frac{1}{4\pi R}\sum_{r=1}^{\infty}(-1)^r\frac{(\ln \abs{u})^{r-1}}{u^r}\sum_{a}\frac{q'_am'_a}{n\cdot p'_a}\Bigg[\left(n\cdot C_1^{\prime(a)}\right)^r\\&~~~~~~~~~~~~~~+\sum_{k=1}^{r-1}\frac{(r-1)!}{(k-1)!(r-k-1)!}n\cdot C_{k+1}^{\prime(a)}\left(\frac{n\cdot p'_a}{m'_a}\right)^k\left(n\cdot C_1^{\prime(a)}\right)^{r-k-1}\Bigg]+\mathcal{O}(\lambda^{-2})~.
\end{split}   
\end{equation}
By expressing the scalar waveform in the observer's frame, we highlight how past events influence present observations. The non-vanishing behavior at infinity signifies that the field's history is embedded in its current state. Soft scalar theorem and scalar memory have been recently derived in \cite{Mao:2017wvx}, see \texttt{\S 2.1 Scalar field}, and \texttt{\S 3.1 Scalar memory} of \cite{Mao:2017wvx}.

\subsection{Scalar waveform for two-body scattering}
\label{subsec3.1}
In this \cref{subsec3.1}, we compute the scalar waveform for two-body scattering. In the $2\to 2$ scattering, we have some simplification, and we finally get  
\begin{equation}\label{20}
\begin{split}
&\phi(u,R\hat{\vb{n}})\simeq-\frac{1}{4\pi R}\Bigg[\frac{q_1m_1}{n\cdot p_1}+\frac{q_2m_2}{n\cdot p_2}-\frac{q'_{1}m'_{1}}{n\cdot p'_{1}}-\frac{q'_{2}m'_{2}}{n\cdot p'_{2}}\\&~~~~~~~~~~~~~~~~~-\left[q_1m_2+q_2m_1+p_1\cdot p_2\left(\frac{q_1}{m_2}\frac{n\cdot p_2}{n\cdot p_1}+\frac{q_2}{m_1}\frac{n\cdot p_1}{n\cdot p_2}\right)\right]\\&~~~~~~~~~~~~~~~~~~~~~~~~~~~\times\sum_{r=1}^{\infty}(-1)^{r}\frac{(\ln u)^{r-1}}{u^r}\left(\sigma^{\textmd{out}}_{12}\right)^{r}\left(\frac{p_1\cdot p_2}{m_1m_2}\right)^{r-1}\left[n\cdot (p_1+p_2)\right]^{r-1}\Bigg]~,
\end{split}
\end{equation}
where $\sigma_{12}^{\textmd{out}}$ the outgoing cross-section is defined as
\begin{equation*}
\sigma_{12}^{\textmd{out}}\equiv\frac{q_1q_2}{4\pi}\frac{m_1^2m_2^2}{\left[(p_1\cdot p_2)^2-m^2_1m^2_2\right]^{\frac{3}{2}}}~,\quad \sigma_{12}^{\textmd{in}}\equiv\frac{q'_1q'_2}{4\pi}\frac{m^{\prime 2}_{1}m^{\prime 2}_{2}}{\big[(p'_{1}\cdot p'_{2})^2-m^{\prime 2}_{1}m^{\prime 2}_{2}\big]^{\frac{3}{2}}}~.
\end{equation*}
The outgoing cross-section \(\sigma_{12}^{\textmd{out}}\) characterizes the probability of interaction between two massive particles exchanging a massless scalar field. 
Since the prefactor \( q_1 q_2 / 4\pi \) is dimensionless, the total mass dimension of \(\sigma_{12}^{\textmd{out}}\) is \(-2\), which is consistent with the expected dimension of a cross-section in natural units (\([\text{length}]^2\) or \([\text{mass}]^{-2}\)). The expression depends on the charges, masses, and momenta of the particles, incorporating relativistic kinematics and phase-space considerations. Similarly, \(\sigma_{12}^{\textmd{in}}\) represents the incoming cross-section. These quantities are essential in studying scattering processes mediated by long-range massless scalar interactions. \vspace{0.2cm}\\
For the incoming waveform, we have to substitute $u\to -u$ and $\ln u\to\ln \abs{u}$
\begin{equation}\label{21}
\begin{split}
\phi(u,R\hat{\vb{n}})&\simeq\frac{1}{4\pi 
 R}\left[q'_1m'_2+q'_2m'_1+p'_1\cdot p'_2\left(\frac{q'_1}{m'_2}\frac{n\cdot p'_2}{n\cdot p'_1}+\frac{q'_2}{m'_1}\frac{n\cdot p'_1}{n\cdot p'_2}\right)\right]\\&~~~~~~~~~~~~~~~~~~~~~~~~~~~~~\times\sum_{r=1}^{\infty}\frac{(\ln \abs{u})^{r-1}}{u^r}\left(\sigma^{\textmd{in}}_{12}\right)^{r}\left(\frac{p'_1\cdot p'_2}{m'_1m'_2}\right)^{r-1}\left[n\cdot (p'_1+p'_2)\right]^{r-1}~.
\end{split}
\end{equation}
Resumming \eqref{20}, we can write down the outgoing waveform as
\begin{equation}
\begin{split}
\phi(u,R\hat{\vb{n}})&\simeq-\frac{1}{4\pi R}\Bigg[\frac{q_1m_1}{n\cdot p_1}+\frac{q_2m_2}{n\cdot p_2}-\frac{q'_{1}m'_{1}}{n\cdot p'_{1}}-\frac{q'_{2}m'_{2}}{n\cdot p'_{2}}\\&~~~~~~~~~~~~+\sigma^{\textmd{out}}_{12}\frac{1}{u}\left[1+\sigma^{\textmd{out}}_{12}\left(\frac{p_1\cdot p_2}{m_1m_2}\right)\left[n\cdot(p_1+p_2)\right]\frac{\ln u}{u}\right]^{-1}D^{\textmd{out}}_{12}\Bigg]~,
\end{split}    
\end{equation}
where $D^{\textmd{out}}_{12}$ is defined by
\begin{equation}
D^{\textmd{out}}_{12}\equiv q_1m_2+q_2m_1+p_1\cdot p_2\left(\frac{q_1}{m_2}\frac{n\cdot p_2}{n\cdot p_1}+\frac{q_2}{m_1}\frac{n\cdot p_1}{n\cdot p_2}\right)~.
\end{equation}
For resumming the incoming waveform \eqref{21}, we have to substitute $u\to -u$ and $\ln u\to\ln \abs{u}$
\begin{equation}
\begin{split}
\phi(u,R\hat{\vb{n}})&\simeq\frac{1}{4\pi R}\sigma^{\textmd{in}}_{12}\frac{1}{u}\left[1-\sigma^{\textmd{in}}_{12}\left(\frac{p'_1\cdot p'_2}{m'_1m'_2}\right)\left[n\cdot(p'_1+p'_2)\right]\frac{\ln \abs{u}}{u}\right]^{-1}D^{\textmd{in}}_{12}~,
\end{split}    
\end{equation}
where $D^{\textmd{in}}_{12}$ is defined by
\begin{equation}
D^{\textmd{in}}_{12}\equiv q'_1m'_2+q'_2m'_1+p'_1\cdot p'_2\left(\frac{q'_1}{m'_2}\frac{n\cdot p'_2}{n\cdot p'_1}+\frac{q'_2}{m'_1}\frac{n\cdot p'_1}{n\cdot p'_2}\right)~.
\end{equation}
The frequency space waveform is defined as
\begin{equation}
\Tilde{\phi}(\omega,R\hat{\vb{n}})\equiv\int du\,\mathrm{e}^{\mathrm{i}\omega u}\phi(u,R\hat{\vb{n}})   
\end{equation}
which is given by
\begin{equation}
\begin{split}
&\Tilde{\phi}(\omega,R\hat{\vb{n}})\\
&\simeq -\frac{\mathrm{i}}{4\pi R}\frac{1}{\omega+\mathrm{i}\epsilon}\Bigg(\frac{q_1m_1}{n\cdot p_1}+\frac{q_2m_2}{n\cdot p_2}-\frac{q'_{1}m'_{1}}{n\cdot p'_{1}}-\frac{q'_{2}m'_{2}}{n\cdot p'_{2}}\Bigg)\\
&~~~~-\frac{\mathrm{i}}{4\pi R}D^{\textmd{out}}_{12}\sum_{r=1}^{\infty}\frac{1}{r!}\omega^{r-1} \left( \ln(\omega+\mathrm{i}\epsilon) \right)^r \left(+\mathrm{i}\sigma^{\textmd{out}}_{12}\right)^{r}\left(\frac{p_1\cdot p_2}{m_1m_2}\right)^{r-1}\left[n\cdot (p_1+p_2)\right]^{r-1}\\&~~~~~\,+\frac{\mathrm{i}}{4\pi R}D^{\textmd{in}}_{12}\sum_{r=1}^{\infty}\frac{1}{r!}\omega^{r-1} \left( \ln(\omega-\mathrm{i}\epsilon) \right)^r \left(-\mathrm{i} {\sigma}^{\textmd{in}}_{12}\right)^{r}\left(\frac{p'_{1}\cdot p'_{2}}{m'_{1}m'_{2}}\right)^{r-1}\left[n\cdot (p'_{1}+p'_{2})\right]^{r-1}~.
\end{split}
\end{equation}
We use the expression for the frequency space waveform 
\begin{equation}
\begin{split}
& \int_{-\infty}^{\infty} \frac{d \omega}{2 \pi} \mathrm{e}^{-\mathrm{i} \omega u} \omega^{k-1}\{\ln (\omega+\mathrm{i} \epsilon)\}^k \simeq \begin{cases}-k!\mathrm{i}^{k-1} \frac{(\ln u)^{k-1}}{u^k} &~\text { for } u \rightarrow+\infty \\
0 &~\text { for } u\rightarrow-\infty\end{cases} \\
& \int_{-\infty}^{\infty} \frac{d \omega}{2 \pi} \mathrm{e}^{-\mathrm{i} \omega u} \omega^{k-1}\{\ln (\omega-\mathrm{i} \epsilon)\}^k \simeq \begin{cases}0 & \text { for } u \rightarrow+\infty \\
+k!\mathrm{i}^{k-1} \frac{(\ln \abs{u})^{k-1}}{u^k} & \text { for } u \rightarrow-\infty~.\end{cases}
\end{split}
\end{equation}
After summing over \( r \), the resummed scalar waveform for two-particle scattering in frequency space\footnote{We thank Carlo Heissenberg and Paolo Di Vecchia for revisiting the derivations in \cref{sec1} and \cref{sec3} and pointing out errors in our earlier expression after this paper appeared on \texttt{arXiv}. We have since re-examined our earlier computations and addressed the identified errors.} is expressed as  
\begin{equation}\label{22}
\begin{split}
&\Tilde{\phi}(\omega,R\hat{\vb{n}})\\&\simeq -\frac{\mathrm{i}}{4\pi R}\frac{1}{\omega+\mathrm{i}\epsilon}\Bigg[\frac{q_1m_1}{n\cdot p_1}+\frac{q_2m_2}{n\cdot p_2}-\frac{q'_{1}m'_{1}}{n\cdot p'_{1}}-\frac{q'_{2}m'_{2}}{n\cdot p'_{2}}\\
&~~~~~~~~~~+\frac{m_1m_2}{(p_1\cdot p_2)[n\cdot (p_1+p_2)]}\Big\{(\omega+\mathrm{i}\epsilon)^{+\mathrm{i}\omega\sigma^{\textmd{out}}_{12}\left(\frac{p_1\cdot p_2}{m_1m_2}\right)[n\cdot(p_1+p_2)]}-1\Big\}D^{\mathrm{out}}_{12}\\&~~~~~~~~~~~~~~~~-\frac{m'_1m'_2}{(p'_1\cdot p'_2)[n\cdot (p'_1+p'_2)]}\Big\{(\omega-\mathrm{i}\epsilon)^{-\mathrm{i}\omega\sigma^{\textmd{in}}_{12}\left(\frac{p'_1\cdot p'_2}{m'_1m'_2}\right)[n\cdot(p'_1+p'_2)]}-1\Big\}D^{\mathrm{in}}_{12}\Bigg]~.
\end{split}
\end{equation}
\subsection{Scalar waveform in purely elastic two-body scattering}\label{subsec3.3}
We assume a purely elastic two-body scattering process in which the following conditions hold
\begin{enumerate}
    \item \textbf{Mass conservation} 
    \[
    m_1 = m'_1 \quad \text{and} \quad m_2 = m'_2~,
    \]

    \item \textbf{Charge conservation} 
    \[
    q_1 = q'_1 \quad \text{and} \quad q_2 = q'_2~.
    \]
\end{enumerate}
The four-momentum is conserved 
\[
p^\mu_1 + p^\mu_2 = p'^\mu_1 + p'^\mu_2~.
\]  
So we have $n\cdot(p_1+p_2)=n\cdot(p'_1+p'_2)$. 
Squaring the momentum conservation equation and using the mass shell conditions \( p_1^2 =-m_1^2 \) and \( p_2^2 =-m_2^2 \) (and similarly for \( p'_1 \) and \( p'_2 \)), we find
\[
p_1 \cdot p_2 = p'_1 \cdot p'_2~.
\]
In this case, we also have $\sigma^{\mathrm{out}}_{12}=\sigma^{\mathrm{in}}_{12}\equiv\sigma_{12}$. The nontrivial cancellation of the memory term occurs if and only if $q_1 = q'_1\,,~m_1 = m'_1$ and $q_2 = q'_2\,,~m_2 = m'_2$. So, the resummed expression of the scalar waveform $\Tilde{\phi}(\omega,R\hat{\vb{n}})$ is given by
\begin{equation}
\label{simwave}
\begin{split}
&\Tilde{\phi}(\omega,R\hat{\vb{n}})\\&\simeq-\frac{\mathrm{i}}{4\pi R}\frac{1}{\omega+\mathrm{i}\epsilon}\frac{m_1m_2}{[n\cdot(p_1+p_2)](p_1\cdot p_2)}\\&~~~~\times\left[(\omega+\mathrm{i}\epsilon)^{+\mathrm{i}\omega\sigma_{12}\left(\frac{p_1\cdot p_2}{m_1m_2}\right)[n\cdot(p_1+p_2)]}D^{\mathrm{out}}_{12}-(\omega-\mathrm{i}\epsilon)^{-\mathrm{i}\omega\sigma_{12}\left(\frac{p_1\cdot p_2}{m_1m_2}\right)[n\cdot(p_1+p_2)]}D^{\mathrm{in}}_{12}\right]\\&\simeq-\frac{\mathrm{i}}{4\pi R}\frac{1}{\omega+\mathrm{i}\epsilon}\bigg[\left(\frac{q_1m_1}{n\cdot p_1}+\frac{q_2m_2}{n\cdot p_2}\right)(\omega+\mathrm{i}\epsilon)^{+\mathrm{i}\omega\mathrm{\Theta}_{12}}-\left(\frac{q_1m_1}{n\cdot p'_1}+\frac{q_2m_2}{n\cdot p'_2}\right)(\omega-\mathrm{i}\epsilon)^{-\mathrm{i}\omega\mathrm{\Theta}_{12}}\\&~~~~~~~~~+\frac{1}{E_{12}\zeta_{12}}\Big(q_1\big(m_2-m_1\zeta_{12}\big)+q_2\big(m_1-m_2\zeta_{12}\big)\Big)\Big((\omega+\mathrm{i}\epsilon)^{+\mathrm{i}\omega\mathrm{\Theta}_{12}}-(\omega-\mathrm{i}\epsilon)^{-\mathrm{i}\omega\mathrm{\Theta}_{12}}\Big)\bigg],
\end{split}
\end{equation}
where we have substituted the expressions of $D_{12}^{\textmd{out/in}}$. Here, $E_{12},\,\zeta_{12}$ and $\mathrm{\Theta}_{12}$ are defined as the following
\begin{equation}
E_{12}\equiv n\cdot(p_1+p_2)~,\quad\zeta_{12}\equiv\frac{p_1\cdot p_2}{m_1m_2}~,\quad\mathrm{\Theta}_{12}\equiv \sigma_{12}\left(\frac{p_1\cdot p_2}{m_1m_2}\right)[n\cdot(p_1+p_2)]~.  
\end{equation}
In the frequency domain, we get
\begin{equation}
\begin{split}
&\Tilde{\phi}(\omega,R\hat{\vb{n}})\simeq-\frac{\mathrm{i}}{4\pi R}\frac{1}{\omega+\mathrm{i}\epsilon}\Big[\texttt{Weinberg's term}+\texttt{Extra term}\Big]~,\\
\end{split}
\end{equation}
where
\begin{equation}
\begin{split}
&\texttt{Weinberg's term}=\left(\frac{q_1m_1}{n\cdot p_1}+\frac{q_2m_2}{n\cdot p_2}\right)(\omega+\mathrm{i}\epsilon)^{+\mathrm{i}\omega\mathrm{\Theta}_{12}}-\left(\frac{q_1m_1}{n\cdot p'_1}+\frac{q_2m_2}{n\cdot p'_2}\right)(\omega-\mathrm{i}\epsilon)^{-\mathrm{i}\omega\mathrm{\Theta}_{12}}~,\\
&\texttt{Extra term}=\frac{1}{E_{12}\zeta_{12}}\Big[q_1\big(m_2-m_1\zeta_{12}\big)+q_2\big(m_1-m_2\zeta_{12}\big)\Big]\Big[(\omega+\mathrm{i}\epsilon)^{+\mathrm{i}\omega\mathrm{\Theta}_{12}}-(\omega-\mathrm{i}\epsilon)^{-\mathrm{i}\omega\mathrm{\Theta}_{12}}\Big]~.
\end{split}
\end{equation}
It is noted in \cite{Vecchia:talk} that the term labeled as \texttt{Weinberg's term} corresponds to the term obtained by using Weinberg's approach \cite{Weinberg:1965nx}. The term referred to as \texttt{Extra term} arises specifically from the worldline approach. In the case of electromagnetism \cite{Karan:2025ndk}, it is absent. The expression \eqref{simwave} is similar to the steps\footnote{We thank Carlo Heissenberg and Paolo Di Vecchia for pointing out this nontrivial cancellation and sharing their unpublished notes with us after this paper appeared on \texttt{arXiv}.} from \texttt{(2.75)} to \texttt{(2.76)} of \cite{Alessio:2024onn} in the gravitational case and from \texttt{(1.18)} to \texttt{(1.19)} of \cite{Karan:2025ndk} in the electromagnetic case.\vspace{0.2cm}\\
Our study focuses on the scalar waveform in frequency space, both in its standard and resummed forms, for the \(2 \to 2\) scattering process. In this case, the formulas are relatively simple. However, when there are more than two incoming or outgoing particles, the expressions become much more complicated. This is due to the complicated structure of the trajectory coefficient \(C_{r+1}^{(a)\mu}\), making it harder to extend the formula to multi-particle scattering.

\subsection{Scalar waveform in the probe approximation}
\label{subsec3.2}
We start by defining the relevant four-momentum components. The four-momentum of the heavy scatterer at rest is given by
\begin{equation}
    P = \left(M, \mathbf{0} \right)~,
\end{equation}
where $M$ is the mass of the scatterer. For probe particles labeled with $a = 1,2, \dots, N$, their four-momentum is given by
\begin{equation}
    p_a = \frac{m_a}{\sqrt{1 - \beta_a^2}} \left(1, \mathbold{\beta}_a \right)~.
\end{equation}
Here, $m_a$ is the mass of the probe particle, $\mathbold{\beta}_a$ represents its velocity in units of the speed of light, and the prefactor accounts for relativistic energy-momentum scaling via the Lorentz factor $\gamma_a\equiv\frac{1}{\sqrt{1 - \beta_a^2}}$.

\paragraph{Scattering process description.}
We consider a scattering process in which $N$ probe particles, each characterized by charge $q_a$, mass $m_a$, and outgoing momentum $p_a$, interact with a heavy scatterer of mass $M$ and charge $Q$ that remains at rest. The probe limit is defined by the conditions
\begin{equation}
    M \gg m_a~, \quad Q \gg q_a~, \quad \frac{Q}{M} \ll \frac{q_a}{m_a}~, \quad \forall a = 1,2, \dots, N~.
\end{equation}

\paragraph{Significance of the probe limit.}
The constraint
\begin{equation}
    \frac{Q}{M} \ll \frac{q_a}{m_a}
\end{equation}
and the condition that the heavy scatterer remains stationary under the scalar influence of the probe particles allow us to neglect its scalar radiation, significantly simplifying the analysis of the scattering process.\vspace{0.2cm}\\
The scalar waveform\footnote{For negative $u$, the memory term will be absent in the waveform with $u\to-u,~\ln u\to\ln\abs{u}$ and $q_a\to q'_a,~m_a\to m_a',~\boldsymbol{\beta}_a\to\boldsymbol{\beta}'_a$.} for positive $u$ under probe approximation is given by
\begin{equation}
\begin{split}
\phi(u,R\hat{\vb{n}})&\simeq\frac{1}{4\pi R}\sum_{a=1}^Nq_a\left\{\frac{1}{\gamma_a\big(1-\hat{\vb{n}}\cdot\mathbold{\beta}_a\big)}-\frac{1}{\gamma'_{a}\big(1-\hat{\vb{n}}\cdot\mathbold{\beta}'_a\big)}\right\}\\&~~~~~~~~~~-\frac{1}{4\pi R}\sum_{r=1}^{\infty}(-1)^{r}\frac{(\ln u)^{r-1}}{u^r} \sum_{a=1}^N\left[\frac{q_aQ}{4\pi}\frac{\big(1-\beta_a^2\big)}{m_a\beta_a^3}\right]^r\frac{q_a\big(\hat{\vb{n}}\cdot\mathbold{\beta}_a\big)}{\gamma_a\big(1-\hat{\vb{n}}\cdot\mathbold{\beta}_a\big)}~.
\end{split}
\end{equation}
We calculate this considering the particle \(a\) as the probe that interacts with the particle \(b\), which acts as a heavy scatterer. In addition, we have shown that the scalar waveform does not depend on the mass $M$ of the scatterer. This is what we should expect in the probe approximation limit.
\section{Conclusions and outlook}
\label{conc}
In this paper, we analyze the scalar waveform, the scalar memory term, and the classical soft scalar theorems using the worldline formalism. We determine an infinite set of subleading terms in the late-time and early-time expansions of the waveform, which also depend only on the momenta and charges of the scattered particles. For two-body scattering, we derive a resummed frequency space scalar waveform, along with the resummed waveform at late-time and early-time. We first compute the trajectories of the scattered particles. We determine the trajectory coefficients for two-body scattering and then compute the trajectory in the probe approximation, which simplifies the analysis. Next, we compute the scalar waveform and determine the scalar memory term. We further refine our analysis by focusing on the scalar waveform specifically in the context of two-body scattering and then computed the waveform in the probe approximation to simplify the problem further. \vspace{0.4cm}\\
We also observe that conditions $p_a\cdot C_1^{(b)} = 0$, (for $b\neq a$) are satisfied for the two-body process that represents the newly identified constraints. These conditions arise naturally from the particle trajectories in two-body scattering processes. In particular, this behavior contrasts with what occurs in electromagnetic interactions, where the standard trajectory coefficient follows \( p_a \cdot C_1^{(a)} = 0 \). The key difference here is that, instead of the momentum of the particle \( a \) enforcing a constraint on its own coefficient \( C_1^{(a)} \), it is the momentum of the particle \( b \) that dictates the condition and vice versa. This distinction is quite striking, suggesting a fundamental deviation in the structure of scalar interactions that may have important implications. The inversion of constraints, where the momentum of a particle governs the trajectory coefficient of its scattering partner, marks a striking departure from electromagnetic behavior. This suggests that scalar interactions involve a more intertwined dynamical structure, where the coupling mechanism inherently links the properties of one particle to the motion of the other. Such cross-momentum constraints could have implications for understanding conserved quantities in scalar field theories, offering a new lens to probe fundamental differences between force carriers (e.g., scalars vs. photons). \vspace{0.4cm}\\
Following conjecture \cite{Karan:2025ndk}, the quantum soft factor for the emission of a single soft scalar can be understood as the ratio of the scattering amplitude (\(\mathcal{S}\)-matrix) with one external soft scalar to the scattering amplitude without it. This soft factor is expected to be expressed in terms of the scalar waveform in frequency space. To test this conjecture, one can perform explicit computations at the one-loop and two-loop levels and extend the analysis to arbitrary loop orders. A key goal of such an investigation is to derive the soft-scalar theorem and verify its consistency with the proposed soft factor. However, there are challenges in this approach. Unlike previously studied soft theorems, the massless scalar in this framework acquires a mass at the loop level. This poses a fundamental problem because the soft limit, crucial for formulating soft theorems, cannot be taken in a straightforward manner. Consequently, the asymptotic symmetries associated with soft scalars, which are valid at the classical level, appear to be broken in the quantum regime.\vspace{0.4cm}\\  
A possible resolution to this issue could involve modifying the theory by introducing additional fields. Such an extension might restore an enhanced symmetry, ensuring that the scalar remains massless even at the loop level. If achieved, this would allow for the formulation of a fully consistent quantum soft scalar theorem.

\subsection*{Remarks on the technical and conceptual challenges of gravitational interactions.}

Importantly, our analysis is performed without considering gravitational interactions, and we discuss the technical and conceptual challenges in analyzing the gravitational case in detail. Following the approach\cite{Karan:2025ndk}, it is quite simple to determine the trajectory coefficients for gravitational interactions if only one needs to focus on the matter part of the stress-energy tensor in the equations of motion\cite{unpublished}. Including the gravity part in the stress-energy tensor gives technical challenges in the computation of trajectory coefficients. A primary difficulty arises because the stress-energy tensor is quadratic in the metric perturbation\footnote{The gravity-part of the stress tensor is defined as $T^h_{\mu\nu}\equiv-\frac{1}{8\pi G_{\textmd{N}}}\Big[\sqrt{-g}G_{\mu\nu}\Big]^{(2)}$ up to the quadratic order in metric perturbation.}, making the computation of the metric perturbation itself highly non-trivial. 
We define $\bar{h}_{\mu\nu}$ as
\begin{equation*}
\bar{h}_{\mu\nu}\equiv h_{\mu\nu}-\tfrac{1}{2}\eta_{\mu\nu}h~,\quad \Rightarrow~h_{\mu\nu}\equiv \bar{h}_{\mu\nu}-\tfrac{1}{2}\eta_{\mu\nu}\bar{h}~.
\end{equation*}
The field equations describe the evolution of the metric perturbation \(\bar{h}_{\mu\nu}\) sourced by the stress-energy tensors of matter and gravity. Our approach follows the post-Newtonian multipolar post-Minkowskian (PN-MPM) approach, described in \cite{Cunningham:2024dog}, which gives a systematic method for modeling gravitational wave generation. The solution is obtained perturbatively by expanding the metric as a series, and we proceed as follows
\begin{enumerate}
\item The first-order perturbation \(\bar{h}^{(1)}_{\mu\nu}\) is determined by solving \(\square \bar{h}^{(1)}_{\mu\nu}=-16\pi G_{\textmd{N}}T^X_{\mu\nu}\), where $T^X_{\mu\nu}$ is the matter part of the stress-energy tensor. 
\item This solution is then used to calculate the gravity stress-energy tensor \(T^h_{\mu\nu}\), which acts as a source of the second-order perturbation \(\bar{h}^{(2)}_{\mu\nu}\).
\item The second-order perturbation is obtained by solving \(\square \bar{h}^{(2)}_{\mu\nu}=-16\pi G_{\textmd{N}}T^h_{\mu\nu}\). This iterative process continues to higher orders, refining the metric solution systematically.
\end{enumerate}
The resummed expression for the gravitational waveform in the $2 \to 2$ scattering process was obtained by computing the waveform in the near-probe regime. It is indeed more difficult to analyze classical dynamics in the case of gravity compared to electromagnetism and scalar theories, due to the presence of nonlinear interactions. In \cite{Fucito:2024wlg}, Fucito, Morales and Russo computed the waveform using the Teukolsky equation formalism that effectively incorporates nonlinear effects through resummation. However, extending this approach beyond the probe limit (i.e., the limit in which $m_1 \ll m_2$), appears to be nontrivial.\vspace{0.2cm}\\
The nonlinear gravitational memory waveform for the scattering of two compact objects in General Relativity at leading order in the post-Minkowskian expansion has been computed in \cite{Georgoudis:2025vkk}. The scattering-amplitudes-based representation of the gravitational waveform is utilized, which naturally expresses the nonlinear memory as the contribution arising from soft gravitons that are emitted by the gravitational waves themselves. This approach provides a systematic framework for capturing the memory effects that emerge from the nonlinear interactions of gravitational radiation in the scattering process.\vspace{0.2cm}\\
Now, even if we attempt to approximate the gravity part of the stress-energy tensor using an averaging procedure, it introduces a conceptual issue. As a result, we currently face obstacles in making further progress. We believe that any effort to solve this problem would be of great significance. We describe in great detail the conceptual issue if we try to make any averaging approximations in the stress-energy tensor of the gravity part. 

\subsubsection*{Conceptual issue.}
We suppose that metric perturbation $\bar{h}^{\mu\nu}_{(b)}$ in transverse satisfies the following equations of motion
\begin{equation}
\square \bar{h}^{\mu\nu}_{(b)}=-16\pi G_{\textmd{N}}T^{\mu\nu}_{(b)}~,   
\end{equation}
where $T^{\mu\nu}_{(b)}$ is the total stress-energy tensor that contains the matter part and the gravity part. The matter part is the following
\begin{equation}
T^{X\mu\nu}_{(b)}(x)=m_b\int d\tau_b\,\delta^{(4)}\big(x-X_b(\tau_b)\big)\frac{dX^\mu_b(\tau_b)}{d\tau_b} \frac{dX^\nu_b(\tau_b)}{d\tau_b}~.    
\end{equation}
The gravity part, in general, is very complicated. We assume that, after applying certain averaging approximations, the gravity part can be represented in the following manner
\begin{equation}\label{100}
T^{h\mu\nu}_{(b)}(x)=\int d\tau_b\,\delta^{(4)}\big(x-X_b(\tau_b)\big)t_b^{\mu\nu}\Bigg(X_b(\tau_b),\frac{dX_b(\tau_b)}{d\tau_b}\Bigg)~,    
\end{equation}
where $t_b^{\mu\nu}$ is symmetric and it depends on $X^\mu_b(\tau_b)$ and $\frac{dX^\mu_b(\tau_b)}{d\tau_b}$. The total stress-energy tensor $T^{\mu\nu}_{(b)}(x)$ must be conserved which means
\begin{equation}
\partial_\mu T^{\mu\nu}_{(b)}(x)=0~.
\end{equation}
It yields the following equation
\begin{equation}
m_b\frac{d^2X^\mu_b(\tau_b)}{d\tau_b^2}+\pdv{\tau_b}\Bigg[\frac{\big(x-X_b(\tau_b) \big)_\nu t_b^{\mu\nu}}{\big(x-X_{b}(\tau_b)\big)\cdot\frac{dX_b(\tau_b)}{d\tau_b}}\Bigg]=0~.
\end{equation}
From the above equation, it is clear that there is no non-trivial $t^{\mu\nu}_b$ that satisfies the above equation except $t^{\mu\nu}_b=-m_b\frac{dX^\mu_b(\tau_b)}{d\tau_b}\frac{dX^\nu_b(\tau_b)}{d\tau_b}$. So, we conclude that the gravity stress-energy tensor cannot be written as \eqref{100}.\vspace{0.4cm}\\ 
The gravity part of the stress-energy tensor cannot be expressed as a localized source tied to particle trajectories \big (e.g. \( T^{h\mu\nu}_{(b)}(x) \propto \int d\tau_b\,\delta^{(4)}(x - X_b) t_b^{\mu\nu} \)\big) without violating conservation. The resulting equations demand \( t^{\mu\nu}_b = -m_b \dot{X}^\mu_b \dot{X}^\nu_b \), which would completely erase the gravity contribution, contradicting its physical necessity. This contradiction implies that such approximations are unphysical, as they violate conservation unless gravity’s contribution is negated, which is nonsensical. \vspace{0.4cm}\\
\noindent Now, we discuss potential future directions. 
\paragraph{Future directions.}
\paragraph{1. Hidden symmetry in scalar interactions from cross-momentum constraints.}
The finding of the cross-momentum constraints \( p_a \cdot C_1^{(b)} = 0 \)\,(for $b\neq a$), which naturally emerge from two-body scattering dynamics in scalar interactions, opens several avenues for further investigation. These constraints show that the momentum of one particle influences the trajectory coefficient of the other particle, which is different from electromagnetic interactions. In electromagnetism, the momentum of a particle affects its own coefficient (\( p_a \cdot C_1^{(a)} = 0 \)). 
\vspace{0.4cm}\\
Could this indicate a hidden symmetry in scalar interactions? Studying these constraints may improve our understanding of how force carrier spin and coupling mechanisms shape the fundamental rules of particle interactions.

\paragraph{2. Toward a consistent quantum soft scalar theorem: resolving loop-induced mass and symmetry breaking.}
We should test the proposed quantum soft factor through loop-level computations and establish a consistent soft scalar theorem. A key challenge is the scalar acquiring mass at the loop level, which disrupts the soft limit and breaks asymptotic symmetries. A possible solution involves modifying the theory by introducing additional fields to restore symmetry and ensure that the scalar remains massless.


 \section*{Acknowledgements}
We thank Carlo Heissenberg and Paolo Di Vecchia for giving helpful comments, and generously sharing their unpublished notes with us after this paper appeared on \texttt{arXiv}. We thank them for revisiting the derivations in \cref{sec1} and \cref{sec3} and pointing out errors in our earlier expression for the $2 \to 2$ resummed frequency space scalar waveform. We reviewed our calculations in \texttt{arXiv [v1]} and corrected the errors. The research of SD is supported by the Shuimu Tsinghua Scholar Program award fellowship from Tsinghua University. The research of PR is supported by the Department of Atomic Energy, Government of India. 

\appendix
\section{Derivation of the particle equations of motion}
\label{pfdvv}
In this appendix, we will derive the particle equations of motion from the action where a massless scalar field is coupled with particles of mass $m_a$ moving along the worldline $X^\mu_a(\tau_a)$. The full action of the classical theory is given by
\begin{equation}\label{12}
S[\phi,X]=\int d^4x\left[-\frac{1}{2}\partial_\mu\phi\partial^\mu\phi+J\phi\right]+\frac{1}{2}\sum_{a} m_a\int d\tau_a \, \left(\frac{dX_a}{d\tau_a}\right)^2~,   
\end{equation}
where $J(x)$ is some source that couples with the field $\phi(x)$. The typical form of the source $J(x)$ is given by
\begin{equation}
J(x)=\sum_a q_a\int d\tau_a\,\delta^{(4)}\big(x-X_a(\tau_a)\big)~,  
\end{equation}
where $q_a$ is the scalar charge of the particle $a$ located at the point $X^\mu_a(\tau_a)$. There are a few things that need to be mentioned. Here, $\tau_a$ must be some parameter associated with the worldline of the particle labeled by $a$. The matter part of the above worldline action is not invariant under reparameterization $\tau_a\to \tau'_a=f_a(\tau_a)$. In addition, the scalar source $J(x)$ is not reparametrization invariant. But in the case of electrodynamics, the four current\footnote{It is $J^\mu(x)=\sum_{a}q_a\int d\tau_a\,\delta^{(4)}\big(x-X_a(\tau_a)\big)\frac{dX_a^\mu(\tau_a)}{d\tau_a}$.} is reparametrization invariant which allows us to choose $\tau_a$ to be an affine parameter\footnote{The matter part of the action in this case would be $-\sum_a m_a\int d\tau_a\,\sqrt{-\dot{X}^2_a}$ which is different what we have taken. But both terms eventually give us the same variation with respect to $X^\mu_a$.} with the normalization $\left(\frac{dX_a}{d\tau_a}\right)^2=-1$. It is also consistent with the Lorentz force law of electrodynamics. But in our case, this is not true. \vspace{0.2cm}\\
Varying the action \eqref{12} with respect to $\phi$, we get
\begin{equation}
\square \phi(x)=-J(x)~.    
\end{equation}
The condition for the action to be invariant under the global shift symmetry $\phi(x) \to \phi(x) + \mathrm{\Lambda}$ is the following
\begin{equation}
\sum_{a} q_a = 0~.
\end{equation}
It is nothing but the conservation of the scalar charge. We consider the following variation of the worldline term with respect to $X^\mu_{a}$. Substituting the expression for $J(x)$ in the action \eqref{12} and varying it with respect to $X^\mu_a$, we can see that the equations of motion are given by
\begin{equation}
m_a\frac{d^2X^\mu_a(\tau_a)}{d\tau_a^2}=q_a\partial^\mu \phi(X_a(\tau_a))~.
\end{equation}
Now consider that the individual source $q_b$ produces the field $\phi_{(b)}(x)$
where $\phi(x)\equiv\sum_{b}\phi_{(b)}(x)$
and we have the complete set of equations of motion written in \eqref{eom}. Also, we have $\phi(X_a(\tau_a))\equiv\sum_{b\neq a}\phi_{(b)}(X_a(\tau_a))$ meaning we have neglected the contribution of $\phi_{(a)}(x)$ at $X^\mu_a(\tau_a)$. Physically, this simply implies that the contribution of self-force is neglected.

\providecommand{\href}[2]{#2}\begingroup\raggedright\endgroup

\end{document}